\pdfoutput=1
\documentclass[preprints,review,accept,moreauthors,pdftex,10pt,a4paper]{mdpi}

\firstpage{1} 
\pubvolume{xx}
\issuenum{1}
\articlenumber{1}
\pubyear{2018}
\copyrightyear{2018}
\externaleditor{Academic Editor: name}
\history{Received: date; Accepted: date; Published: date}
\usepackage{bm}

\newcommand{\sub}[1]{$_{\mathrm {#1}}$}
\newcommand{\subm}[1]{_{\mathrm {#1}}}
\newcommand{\sps}[1]{$^{\mathrm {#1}}$}

\newcommand{\etal}{\textit{et~al.}}

\newcommand{\Tc}{T\subm{c}}

\newcommand{\x}{{\boldmath$X$}}

\newcommand{\Hcc}{H\subm{c2}}

\newcommand{\BS}{Bi$_2$Se$_3$}

\newcommand{\cbs}{Cu$_{x}$Bi$_2$Se$_3$}
\newcommand{\sbs}{Sr$_{x}$Bi$_2$Se$_3$}
\newcommand{\nbs}{Nb$_{x}$Bi$_2$Se$_3$}
\newcommand{\abs}{$A_{x}$Bi$_2$Se$_3$}

\newcommand{\bs}{Bi$_2$Se$_3$}
\newcommand{\bt}{Bi$_2$Te$_3$}
\newcommand{\Dx}{\Delta_{4x}}
\newcommand{\Dy}{\Delta_{4y}}
\newcommand{\sro}{Sr$_2$RuO$_4$}

\newcommand{\cxpsbs}{Cu$_{x}$(PbSe)$_5$(Bi$_2$Se$_3$)$_6$}
\newcommand{\psbs}{(PbSe)$_5$(Bi$_2$Se$_3$)$_6$}

\newcommand{\rev}[1]{#1}
\newcommand{\revv}[1]{#1}

\Title{Nematic superconductivity\\in doped Bi$_2$Se$_3$ topological superconductors}

\Author{Shingo Yonezawa $^{1}$\orcidA{}}

\AuthorNames{Shingo Yonezawa}

\address{%
$^{1}$ \quad Department of Physics, Graduate School of Science, Kyoto University, Japan; yonezawa@scphys.kyoto-u.ac.jp}

\corres{Correspondence: yonezawa@scphys.kyoto-u.ac.jp; Tel.: +81-75-753-3744}

\abstract{Nematic superconductivity is a novel class of superconductivity characterized by spontaneous rotational-symmetry breaking in the superconducting gap amplitude and/or Cooper-pair spins with respect to the underlying lattice symmetry. 
Doped Bi$_2$Se$_3$ superconductors, such as \cbs, \sbs, and \nbs, are considered as candidates for nematic superconductors, in addition to the anticipated topological superconductivity. 
\revv{Recently}, various bulk probes, such as nuclear magnetic resonance, specific heat, magnetotransport, magnetic torque, and magnetization,  have consistently revealed two-fold symmetric behavior in their in-plane magnetic-field-direction dependence, although the underlying crystal lattice possesses three-fold rotational symmetry. 
More recently, nematic superconductivity is directly visualized using scanning tunneling microscopy and spectroscopy.
In this short review, \revv{we} summarize the current researches on the nematic behavior in superconducting doped Bi$_2$Se$_3$ systems, and discuss issues and perspectives.}


\begin{document}
%%%%%%%%%%%%%%%%%%%%%%%%%%%%%%%%%%%%%%%%%%

\section{Introduction}

% Topo materials, Topo-SC, Majorana, 

In the last decade, the research field on topological materials, which possess non-trivial topology in their electronic-state wave functions in the reciprocal space, has been expanding substantially~\cite{Hasan2010.RevModPhys.82.3045, Qi2011.RevModPhys.83.1057, Ando2013.JPhysSocJpn.82.102001}. 
As a counterpart of topological insulators, it has been recognized that \revv{certain superconductors can have non-trivial topological nature in their wavefunctions}~\cite{Qi2011.RevModPhys.83.1057, Schnyder2015.JPhysCondensMatter.27.243201, Sato2017.RepProgPhys.80.076501}.
Superconductivity with such non-trivial wavefunction topology is now called topological superconductivity.
It has been predicted that topological superconductivity can lead to various novel phenomena.
In particular, the Majorana quasiparticles with non-Abelian zero-energy modes, hosted in topologically protected edge states or vortex cores, are quite intriguing~\cite{Wilczek2009.NaturePhys.5.614}.
The realization and utilization of such Majorana modes are one of the holy grails of this research field.

% Bulk Topo-SC
%One promising route to realize Majorana modes is to use superconductor-based heterostructures and add an explicit time-reversal symmetry breaking such as applied magnetic field or spontaneous magnetism in a material.
%One class of such superconductors is topological materials with topological surface state exhibiting conventional $s$-wave superconductivity in its bulk.
%In such a material, the topological surface state with locked spin degree of freedom exhibit superconductivity due to proximity effect from the bulk and this superconductivity can have non-trivial topological nature inherited from the unusual spin-orbital locking of the surface state.
%Examples of such a class of topological superconductors, which shall be called as self-proximity topological superconductors, are Fe(Se,Te) and Tl$_x$Bi$_2$Se$_3$.

There are a number of superconductors that seemingly exhibit topological superconductivity in bulk~\cite{Yonezawa2016.AAPPSBulletin.26.3}.
Such bulk topological superconductors are the prototype of topological superconductivity, although there are now other recipes to induce topological superconductivity by making use of the proximity effect.
Recently, a candidate for a bulk topological superconductor, doped \bs, has been found to exhibit unusual rotational-symmetry breaking in the superconducting (SC) gap {\em amplitude} as well as in the SC {\em spin} degree of freedom~\cite{Matano2016.NaturePhys.12.852, Yonezawa2017.NaturePhys.13.123, Pan2016.SciRep.6.28632, Asaba2017.PhysRevX.7.011009}.
This phenomena, called as ``nematic'' superconductivity~\cite{Fu2014.PhysRevB.90.100509}, has been attracting much attention as a new species in the superconductor zoo, accompanied by a novel class of symmetry breaking in its SC wave function.
\revv{The {\em shape} and {\em topology} of the SC wave function are closely related.
Indeed, in \abs, it has been theoretically known that the nematic superconductivity is accompanied by non-trivial topological SC gap~\cite{Fu2010.PhysRevLett.105.097001, Ando2015.AnnuRevCondensMatterPhys.6.361, Sasaki2015.PhysicaC.514.206, Sato2017.RepProgPhys.80.076501}. Therefore, these new experimental observations of the nematic SC gap and spin have been providing firm bulk evidence for topological superconductivity, and thus establish strong bases toward realization and manipulation of Majorana states in this class of compounds.} 

In this short review, \revv{we} summarize recent observations of the nematic behavior in superconducting doped \bs\ systems.
After a brief introduction of nematic superconductivity in Sec.~\ref{sec:nematic-SC}, \revv{we} explain experimental and theoretical understandings of superconductivity in \abs\ in Sec.~\ref{sec:doped-Bi2Se3}.
Section~\ref{sec:nematic-experiment} is devoted to explain recent experimental findings of nematic superconductivity.
Then \revv{we} discuss several known issues in Sec.~\ref{sec:issues}, before summarizing the content in Sec.~\ref{sec:conclusion}.

\section{Nematic Superconductivity: Rotational Symmetry Breaking in the Gap Amplitude}
\label{sec:nematic-SC}

\subsection{Symmetry Breaking in Superconductivity}

\begin{figure}[tb]
\centering
\includegraphics[width=1\textwidth]{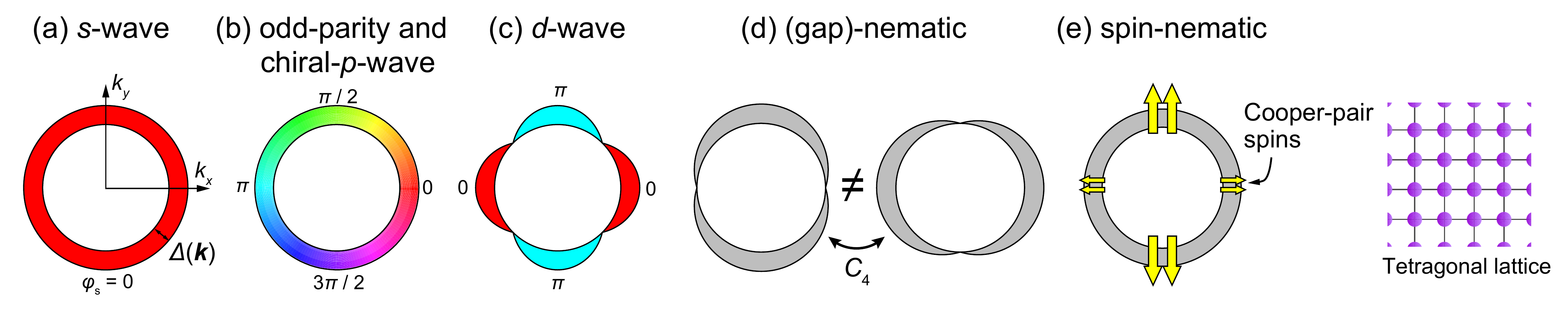}
\caption{Schematic comparison of various known superconductivity and gap/spin nematic superconductivity, for the case of a tetragonal-lattice system.}
\label{fig:nematic-schematic}
\end{figure}

The concept of the spontaneous symmetry breaking has fundamental importance in superconductivity. 
In the Bardeen-Cooper-Schrieffer (BCS) theory~\cite{Bardeen1957.Phys.Rev.108.1175}, the SC state spontaneously breaks the $U(1)$ gauge symmetry, even for the ordinary $s$-wave superconductivity (Fig.~\ref{fig:nematic-schematic}(a)).
It has been then an interesting and long-standing question whether SC states with additional symmetry breaking exists or not.
Such superconductivity with additional symmetry breaking is called unconventional superconductivity. 

For example, superconductivity with spontaneous time-reversal symmetry breaking in the orbital part of the SC order parameter (Fig.~\ref{fig:nematic-schematic}(b)) is called as ``chiral'' superconductivity, and is believed to be realized in several materials such as \sro~\cite{Luke1998.Nature.394.558,Xia2006.PhysRevLett.97.167002}, URu\sub{2}Si\sub{2}~\cite{Yamashita2014.NaturePhys.11.17,Schemm2015.PhysRevB.91.140506} and UPt\sub{3}~\cite{Schemm2014.Science.345.190}.
Odd parity superconductivity, which exhibit $\pi$ phase shift under spacial inversion (see again Fig.~\ref{fig:nematic-schematic}(b)), has been an interesting topic for fairly a long time~\cite{Anderson1961.PhysRev.123.1911, Balian1963}.
Odd-parity superconductivity (and superfluidity) is confirmed in superfluid $^{3}$He~\cite{Leggett1975.RevModPhys.47.331, Mizushima2016.JPhysSocJpn.85.022001}, and is very probably realized in \sro~\cite{Mackenzie2003RMP, Maeno2012.JPhysSocJpn.81.011009, Mackenzie2017.npjQuantumMater.2.40} and UPt\sub{3}~\cite{Izawa2014.JPhysSocJpn.83.061013}. 
Moreover, odd-parity superconductivity is now recognized as an key ingredient to realize topological superconductivity~\cite{Sato2010.PhysRevB.81.220504, Fu2010.PhysRevLett.105.097001}.

Another fundamentally important symmetry is rotational symmetry.
The symmetry behavior under rotation provides a basis of the classification of superconductivity into $s$-wave, $p$-wave, $d$-wave, $f$-wave, etc. 
However, the infinitesimal rotational symmetry $C_\infty$ is already broken in superconductors because of the crystal lattice. 
Thus, in reality, what actually matters is the breaking / invariance of the discrete $C_n$ rotational symmetry of the underlying lattice.
For example, in $d$-wave superconductivity in a tetragonal system (Fig.~\ref{fig:nematic-schematic}(c)), the phase factor for one $k$ direction and for its perpendicular direction differs by $\pi$. 
Therefore, the four-fold rotation $C_4$ symmetry is broken in the phase factor of the SC order parameter.

% d波とかとの比較、Gap Nematic/Spin Nematic
\subsection{Gap-Nematic and Spin-Nematic Superconductivity}

Such a rotational symmetry breaking in the {\em phase} degree of freedom in non-$s$-wave superconductivity is intriguing but the experimental detection of such symmetry breaking is actually very difficult. 
This is because the SC gap amplitude, which governs most of superconducting properties, is actually {\em invariant} under the $C_n$ rotation.
To detect the rotational symmetry breaking in the phase factor, one has to utilize sophisticated interference techniques, as performed in cuprate $d$-wave superconductors~\cite{vanHarlinen.RevModPhys.67.515, Tsuei2000.RevModPhys.72.969}.
In contrast, if the rotational-symmetry breaking occurs in the SC gap {\em amplitude} as shown in Fig.~\ref{fig:nematic-schematic}(d), the rotational symmetry breaking would be more robust and be detectable in principle in any bulk quantities. 
Such superconductivity with broken rotational symmetry in the gap {\em amplitude}, was named as ``nematic superconductivity'' first by Fu~\cite{Fu2014.PhysRevB.90.100509}, and has been attracting much attention as a new class of superconductivity accompanied by a novel spontaneous symmetry breaking.

% 通常のelectron nematicityとの比較

The word ``nematic'', originally used in the research field of liquid crystals, refers to \revv{the states with} the spontaneous rotational symmetry breaking of the bar-shaped liquid crystal molecules but without breaking the translational symmetry.
Now this word is imported to solid-state physics, and the ``nematic electron liquid'' with spontaneous rotational symmetry breaking in the conduction electron without losing conductivity has been attracting much attention and actually found in various systems such as cuprates~\cite{Ando2002.PhysRevLett.88.137005, Vojta2009.AdvPhys.58.699}, iron pnictides~\cite{Kasahara2012.Nature.486.382, Fernandes2014.NaturePhys.10.97}, and a ruthenate~\cite{Borzi2007.Science.315.214}. 
The nematic superconductivity is a superconducting version of such nematic electron liquids, but occurring as a consequence of Cooper-pair formation.

In the case of spin-triplet superconductivity, the spin part of the order parameter, i.e. the $\bm{d}$ vector, can also exhibit rotational symmetry breaking.
Here, we should be careful that there are two different levels of the spin-rotational-symmetry breaking.
Firstly, the $SO(3)$ symmetry of the spin space breaks in the presence of spin-orbit interaction.
In this case, the spin susceptibility exhibit anisotropic behavior but still obeys the symmetry of the lattice.
This first kind of the spin-rotational symmetry breaking has not been observed in a leading candidate spin-triplet superconductor \sro~\cite{Murakawa2004.PhysRevLett.93.167004} and was weakly observed in another candidate UPt\sub{3}~\cite{Tou1998.PhysRevLett.80.3129}.
A more exotic phenomenon is that the spin part even breaks the $C_n$ lattice rotational symmetry (but without spin polarization).
In this case, spin susceptibility exhibit rotational-symmetry breaking, as schematically shown in Fig.~\ref{fig:nematic-schematic}(e).
\revv{We} shall call this phenomenon as ``spin-nematic superconductivity''; and for distinguishment, the nematic superconductivity with the rotational symmetry breaking in the gap amplitude is also called as ``gap-nematic superconductivity''.
Spin-nematic superconductivity has never been known in any material before its discovery in \cbs~\cite{Matano2016.NaturePhys.12.852}.

%%%%%%%%%%%%%%%%%%%%%%%%%%%%%%%%%%%%%%%%%%
\section{Superconductivity in Doped Bi$_2$Se$_3$}
\label{sec:doped-Bi2Se3}

In this section, SC and normal-state properties of doped \bs, as well as theoretically proposed SC states, are described.
For more detailed information, refer to the nice reviews found in \revv{Refs.~\cite{Ando2015.AnnuRevCondensMatterPhys.6.361, Sasaki2015.PhysicaC.514.206, Sato2017.RepProgPhys.80.076501}}.

\subsection{Crystal Structure of the Mother Compound Bi$_2$Se$_3$}

\begin{figure}[tb]
\centering
\includegraphics[width=0.8\textwidth]{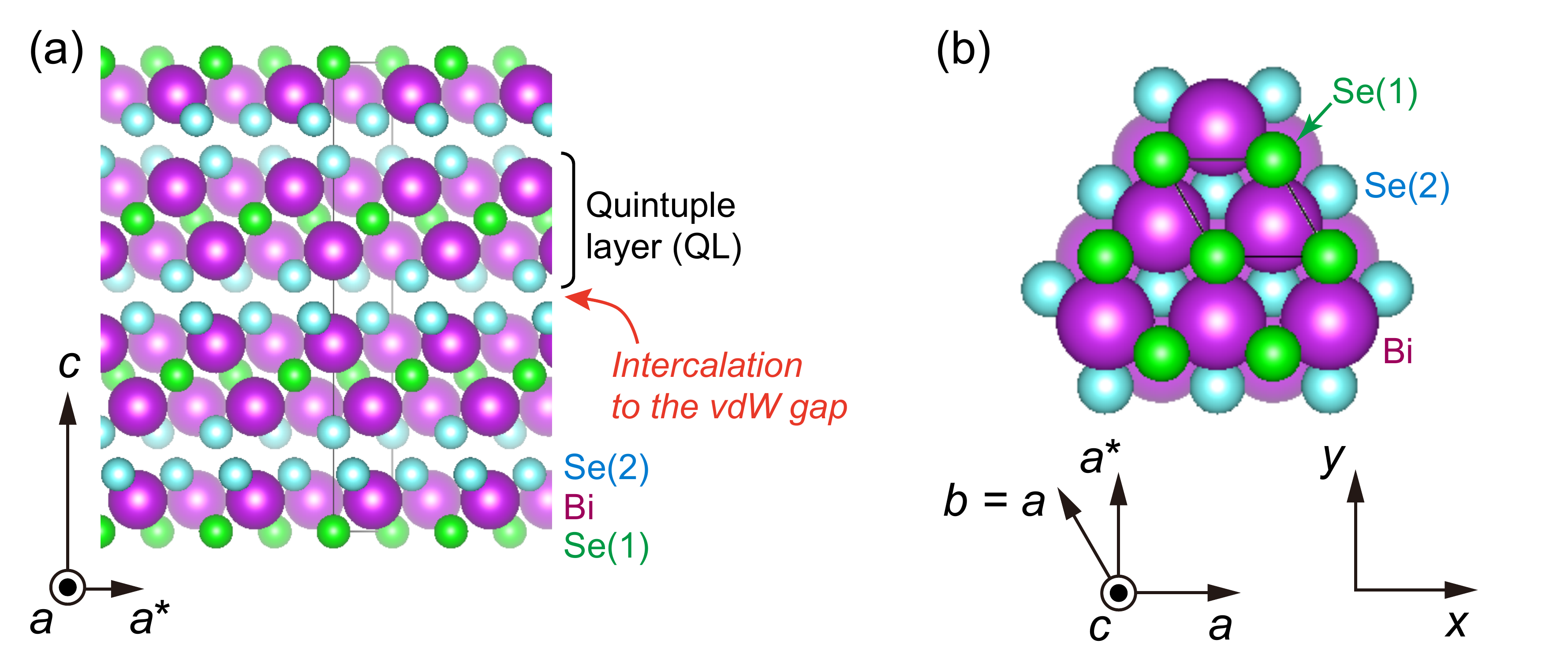}
\caption{Schematic description of the crystal structure of the mother compound \bs~\cite{Nakajima1963.JPhysChemSolids}. The purple spheres are the Bi atoms, and the green and light-blue spheres are the Se(1) and (2) atoms, respectively. The color of the spheres are modified depending on the depth along the view direction: Atoms closer to the view point have thicker colors. (a) View from the $a$-axis direction. The intercalated metallic ions most likely sit in the van der Waals (vdW) gap between the quintuple layers (QL). (b) View from the $c$-axis direction. The crystal structure figures were made using the software VESTA-3~\cite{Momma2011.JApplCrytallogr.44.1272}.}
\label{fig:crystal-structure}
\end{figure}

\bs, the mother compound of the doped \bs\ superconductors, has been extensively studied as a prototypical topological insulator~\cite{Zhang2009.NaturePhys.5.438, Xia2009.NaturePhys.5.398}.
This compound has a rhombohedral (or trigonal) crystal structure as shown in Fig.~\ref{fig:crystal-structure}, with the space group of $R\bar{3}m$ ($D^5_{3d}$)~\cite{Nakajima1963.JPhysChemSolids}. 
The crystal structure contains three equivalent $a$ axes.
Strictly speaking, the situations with $H$ parallel to the $a$ axis and $-a$ axis are different because of the trigonal symmetry and the pseudo-vector nature of the magnetic field.
Nevertheless, practically, $H\parallel a$ and $H\parallel -a$ are almost equivalent in most cases and thus many physical quantities are expected to exhibit pseudo-six-fold rotational symmetry as a function of in-plane field direction.
The $a^\ast$ axis is perpendicular to the $a$ axis within the $ab$ plane.
Importantly, the $a^\ast$-$c$ plane is a mirror-symmetry plane, whereas the $a$-$c$ plane is not, as clearly seen in Fig.~\ref{fig:crystal-structure}(b).
This existence/absence of the mirror symmetry is closely related to the stability of the gap nodes in the nematic SC state, as will be discussed in Sec.~\ref{subsec:sc-states}.
Throughout this paper, we define the $x$ axis along one of the three $a$ axes. 
In most cases, we choose $x$ to be along the ``special'' $a$ axis due to the nematic SC order (e.g. the $a$ axis with maximal or minimal $\Hcc$). 
Then the $y$ axis is defined so that it is perpendicular to the $x$ axis within the $ab$ plane, as explained in Fig.~\ref{fig:crystal-structure}(b).

The crystal structure consists of Se-Bi-Se-Bi-Se layers, as shown in Fig.~\ref{fig:crystal-structure}(a).
This set of layers is called as the quintuple layer (QL). 
Between the QLs, there is a van-der Waals (vdW) gap, through which metallic ions penetrate the sample during the synthesis process.
Thus the ions are most likely to be intercalated into a certain site within the vdW gap.
For Sr-doped \bs, Sr ions may also sit in a interstitial site~\cite{Li2018.PhysRevMaterials.2.014201}.
The precise position of the doped ions, in particular that for superconducting samples, has not been fully clarified.

\subsection{Basic Properties of Doped Bi$_2$Se$_3$ Superconductors}

% SCの発見, Cu-BiSe, Sr and Nb doping
In 2010, the pioneering work by Hor \etal\ revealed that \cbs\ exhibits superconductivity below the critical temperature $\Tc$ of around 3~K~\cite{Hor2010.PhysRevLett.104.057001}. 
This is truly the beginning of the research field of superconducting doped \bs\ systems, but the initial samples grown by ordinary melt-grown technique exhibited the superconducting \revv{shielding} fraction of \revv{about 20\%} and did not show complete vanishing of resistivity.
One year later, Kriener \etal\ found that, with the electrochemical intercalation of Cu and with a suitable annealing process, \cbs\ indeed exhibit bulk superconductivity with clear zero resisivity and volume fractions reaching 60-70\% evaluated from specific-heat and magnetization measurements~\cite{Kriener2011.PhysRevLett.106.127004, Kriener2011.PhysRevB.84.054513}.
In 2015, Sr doping or Nb doping were also found to drive \bs\ to superconduct, with $\Tc$ again around 3~K~\cite{Liu2015.JAmChemSoc.137.10512, Shruti2015.PhysRevB.92.020506, Qiu2015.arXiv:1512.03519.Full}.
In contrast to the Cu-doped material, \sbs\ and \nbs\ exhibit bulk superconductivity with fairly a large \revv{shielding} fraction close to 100\% even with melt-grown samples. 

These three compounds exhibit similar SC behavior but there are several significant differences as well.
Firstly, the normal-state electronic state is different.
In \cbs, an ellipsoidal or cylindrical Fermi surface, depending on the carrier density, has been revealed by angle-resolved photoemission spectroscopy (ARPES) and quantum oscillation experiments~\cite{Lahoud2013.PhysRevB.88.195107}.
\revv{Superconducting \cbs\ typically has the carrier density $n$ of around $10^{20}$~cm$^{-3}$~\cite{Hor2010.PhysRevLett.104.057001, Kriener2011.PhysRevLett.106.127004, Lahoud2013.PhysRevB.88.195107}.
Furthermore, $n$ is known to be rather insensitive to the Cu concentration~\cite{Kriener2011.PhysRevB.84.054513}.
In \sbs, $n$ tends to be even lower than that in \cbs: $n$ for superconducting \sbs\ is consistently reported to be $\sim 2\times 10^{19}$~cm$^{-3}$~\cite{Liu2015.JAmChemSoc.137.10512, Shruti2015.PhysRevB.92.020506}.}
For \nbs, the quantum oscillation consists of oscillations with \revv{different frequencies and with different field-angular dependence}, indicating that the Fermi surface is not a simple ellipsoid or cylinder but is composed of multiple pockets~\cite{Lawson2016.PhysRevB.94.041114}.
This is crucially different from the single Fermi surface in \cbs\ and \sbs.
In addition, \nbs\ has some controversy on the magnetism: in the initial report, it was argued that \nbs\ exhibits long-range magnetic order in the superconducting state and this magnetic order assists formation of a chiral SC state~\cite{Qiu2015.arXiv:1512.03519.Full}. 
In later reports, however, such magnetism has not been reported~\cite{Asaba2017.PhysRevX.7.011009, Shen2017.npjQuantumMater.2.59}.
Secondly, a practical difference among the three compounds is that \sbs\ and \nbs\ are stable in air, whereas superconductivity in \cbs\ is known to diminish if a sample is kept in air~\revv{\cite{Hor2010.PhysRevLett.104.057001}}.
\revv{We should comment here that single-crystalline \cbs\ prepared by electrochemical intercalation might not be as air-sensitive as melt-grown samples~\cite{Kriener.PrivateCommun}.}
Because of \revv{the higher} stability, Sr and Nb-doped \bs\ have been extensively studied since their discoveries, as reviewed below.

% Normal-state ARPES, effective hamiltonian, ...
The normal-state electronic state of \cbs\ was investigated with ARPES and quantum oscillation experiments~\cite{Lahoud2013.PhysRevB.88.195107}. 
It was found that the electronic band structure of \cbs\ is essentially the same as that of \bs, and that the surface state originating from the topological-insulator nature of \bs\ still exist even after the Cu doping. 
As expected, \cbs\ is heavily electron doped compared to the mother compound: the chemical potential is located 0.2-0.5~eV above the Dirac point of the topological surface state.
Still, the surface state was clearly observed even at the chemical potential, well distinguishable from the bulk band.
This means that the \revv{bulk} conduction electrons on the Fermi surface of \cbs\ inherit the ``twisted'' \revv{nature of the bulk electronic state} of the mother compound.
Once superconductivity sets in, the Cooper pairs are formed among \revv{bulk} electrons in such a non-trivial topological state.
Such situation is favorable for the realization of odd-parity and  topological SC states even for simple pairing interactions, as described in the next subsection.
The preserved topological-insulator surface state after doping was also confirmed in \sbs\ via quantum oscillation and scanning tunneling microscope (STM) experiments~\cite{Liu2015.JAmChemSoc.137.10512, Du2017.NatureCommun.8.14466}

\subsection{Possible Superconducting States}
\label{subsec:sc-states}

Just after the discovery of superconductivity in \cbs~\cite{Hor2010.PhysRevLett.104.057001}, Fu~\etal\ performed theoretical analysis on possible SC states realized in this compound~\cite{Fu2010.PhysRevLett.105.097001}. 
The result is quite surprising since odd-parity topological superconductivity was predicted even with a simple pairing interaction.
Previously, it had been believed that an unconventional pairing glue such as ferromagnetic spin fluctuation is required to realize bulk odd-parity superconductivity.
Very naively, the odd-parity superconductivity in this model originates from strong orbital mixing on the Fermi surface;
when a Cooper pair is formed among electrons in different orbitals, odd-parity superconductivity is rather easily realized.

Let us review the result of this theory in a bit more detail. 
Fu~\etal\ considered the $D_{3d}$ point group of \bs\ and assumed Cooper pairing of electrons in two $p_z$ orbitals localized nearly at the top and bottom of a QL.
On the other hand, the pairing interaction was assumed to be point-like. 
Then, six possible pairing states $\Delta_{1a}$, $\Delta_{1b}$, $\Delta_2$, $\Delta_3$, $\Delta_{4x}$, and $\Delta_{4y}$, were proposed as listed in Table~\ref{table:SC-states}.
Here, the pairing potential in the orbital bases is expressed with Pauli matrices in the orbital and spin spaces, $\sigma_\mu$ and $s_\mu$ ($\mu = x, y, z$);
Thus those with off-diagonal terms in the orbital matrices $\sigma$, i.e., those containing $\sigma_x$ or $\sigma_y$, are inter-orbital pairing states and $\sigma_y$ characterizes an orbital-singlet state~\cite{Sasaki2015.PhysicaC.514.206}.
Among them, $\Delta_{1a}$ and $\Delta_{1b}$ are even-parity states and the others, $\Delta_{2}$, $\Delta_3$, $\Delta_{4x}$ and $\Delta_{4y}$ are the odd-parity states.
All of these odd-parity states belong to topological superconductivity, because odd-parity superconductivity is proved to have a non-trivial topological nature if the Fermi surface encloses an odd number of high-symmetry points in the Brillouin zone~\cite{Sato2010.PhysRevB.81.220504, Fu2010.PhysRevLett.105.097001}.
The states $\Delta_2$, $\Delta_{4x}$ and $\Delta_{4y}$ contain $s$ matrices in the pairing potential and thus are fundamentally spin-triplet SC states even in the absence of the spin-orbit interaction. 
In contrast, $\Delta_3$ in the absence of the spin-orbit interaction is a spin-singlet state in spite of odd-parity nature (Notice that $\bm{d} = 0$ for $\lambda = 0$), since the Pauli principle is satisfied together with the orbital degree of freedom~\cite{Hashimoto2013.JPhysSocJpn.82.044704}.
Nevertheless, with finite spin-orbit interaction, $\Delta_3$ also acquires spin-triplet nature.

To evaluate the SC-gap and $\bm{d}$-vector structures in the reciprocal space, the pair potential in the orbital bases has to be converted to the SC order parameter in the band bases.
In the work by Hashimoto~\etal~\cite{Hashimoto2013.JPhysSocJpn.82.044704, Hashimoto2014.SupercondSciTechnol.27.104002}, the $\bm{d}$ vector for the lowest order in $k$ was evaluated as listed in Table~\ref{table:SC-states}.
Here, the $\bm{d}$ vector depends on the ratio $\lambda$ between the spin-orbit interaction (denoted as $v$ in Ref.~\cite{Hashimoto2013.JPhysSocJpn.82.044704}) and the coefficient $v_z$ describing the $k_z$-linear term in the bulk electronic band dispersion.
These values are estimated to be $v = 4.1$~eV\,\AA\ and $v_z = 9.5$~eV\,\AA~\cite{Hashimoto2013.JPhysSocJpn.82.044704, Hashimoto2014.SupercondSciTechnol.27.104002, Zhang2009.NaturePhys.5.438}; thus $\lambda$ is roughly 0.5.
In addition, there is a small gapping term $\varepsilon$ to express the disappearance of the point nodes in the $\Delta_{4y}$ pairing~\cite{Fu2014.PhysRevB.90.100509}.
Experimentally, this term is expected to be fairly small~\cite{Yonezawa2017.NaturePhys.13.123}.
In the bottom row of Table~\ref{table:SC-states}, the $\bm{d}$ vector structure in the $k$ space, as well as the gap $|\bm{d}(\bm{k})|$, are schematically shown for a spherical Fermi surface and with the parameters $\lambda = 0.5$ and $\varepsilon = 0.1$.
The $\bm{d}$-vector structure of each odd-parity state has a complicated texture on the Fermi surface, being similar to the $\bm{d}$-vector structure in the Balian-Werthamer (BW) state (the B phase of $^{3}$He)~\cite{Balian1963} but quite different from a $k$-uniform $\bm{d}$ vector in, e.g., \sro~\cite{Mackenzie2003RMP, Maeno2012.JPhysSocJpn.81.011009}.

\begin{table}[H]
\caption{Proposed superconducting states for doped \BS~\cite{Fu2010.PhysRevLett.105.097001, Hashimoto2013.JPhysSocJpn.82.044704, Hashimoto2014.SupercondSciTechnol.27.104002, Sasaki2015.PhysicaC.514.206, Ando2015.AnnuRevCondensMatterPhys.6.361}. 
The $\bm{d}$ vector structures in the band bases are from Refs.~\cite{Hashimoto2013.JPhysSocJpn.82.044704, Hashimoto2014.SupercondSciTechnol.27.104002}.
Here, $\lambda$ represents the strength of the spin-orbit coupling; and $\varepsilon$ represents the gap minima for the $\Delta_{4y}$ state (see text).
In the bottom row, schematic gap and $d$-vector structures of each state are shown, together with various cut views. 
The $\lambda$ value is chosen to be 0.5 and the $\varepsilon$ value is just set to be 0.1.
The sphere at a center of the cut views is the Fermi surface. 
The gap structure is expressed with colored surfaces, whose distance from the Fermi surface corresponds to the SC gap amplitude $|\bm{d}|$ normalized by its maximal value $d_0$.
The color of this surface also depicts the gap value, as well as the $d$-vector direction, as explained in the left bottom cell:
The hue and lightness of the color indicate the azimuthal and polar angles of the $d$ vector, $\phi_{\bm{d}}$ and $\theta_{\bm{d}}$, respectively; whereas the grayness of the color depicts the normalized gap, with 50\%-gray corresponding to $|\bm{d}| = 0$.
}
\label{table:SC-states}
\centering
\tablesize{\footnotesize} %% You can specify the fontsize here, e.g.  \tablesize{\footnotesize}. If commented out \small will be used.
\renewcommand{\arraystretch}{1.4}
\begin{tabular}{ccccccc}
\toprule
 & \textbf{\boldmath{$\Delta_{1a}$}} & \textbf{\boldmath{$\Delta_{1b}$}}	& \textbf{\boldmath{$\Delta_{2}$}} & \textbf{\boldmath{$\Delta_{3}$}}	& \textbf{\boldmath{$\Delta_{4x}$}} & \textbf{\boldmath{$\Delta_{4y}$}} \\
 \midrule
%%%
\parbox[c][3em][c]{0.12\textwidth}{\centering Irreducible representation} & $A_{1g}$ &  $A_{1g}$  & $A_{1u}$ & $A_{2u}$ & \multicolumn{2}{c}{$E_u$} \\%
%%%
\parbox[c][3em][c]{0.12\textwidth}{\centering Pairing potential} & $\sigma_0$ & $\sigma_x$ & $\sigma_y s_z$ & $\sigma_z$ & $\sigma_y s_x$ & $\sigma_y s_y$ \\
%%%
$\bm{d}$-vector &  - & - & $\sim(\lambda k_x, \lambda k_y, k_z)$ & $\sim(-\lambda k_y, \lambda k_x, 0)$ & $\sim(k_z, 0, -\lambda k_x)$ & $\sim(\varepsilon k_x, -k_z, \lambda k_y)$ \\
%%%
Parity    & even &  even  & odd & odd & \multicolumn{2}{c}{odd} \\%
%%%
%Spin      & singlet &  singlet  & triplet & singlet & \multicolumn{2}{c}{triplet} \\%
%%%
%Orbital   & triplet &  triplet  & singlet & triplet & \multicolumn{2}{c}{singlet} \\%
%%%
Topo. SC  & no &  no  & yes & yes & \multicolumn{2}{c}{yes} \\%
%%%
Nematic SC  & no &  no  & no & no & \multicolumn{2}{c}{yes} \\%
%%%

\parbox[b][0.5\textwidth][c]{0.12\textwidth}{\centering Schematic $\bm{d}$-vector / gap structures \includegraphics[width=0.12\textwidth]{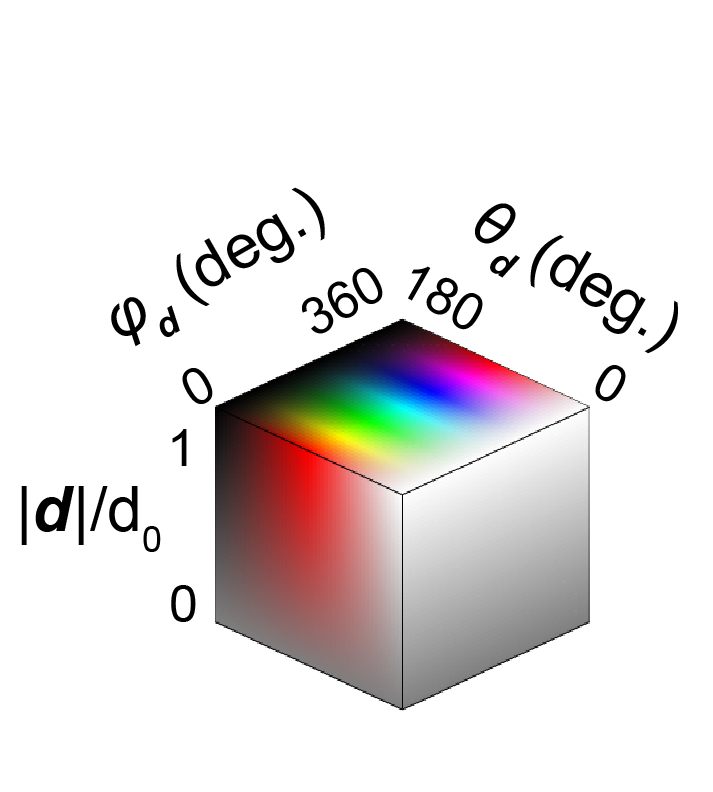}} & \multicolumn{2}{c}{\centering \includegraphics[width=0.14\textwidth]{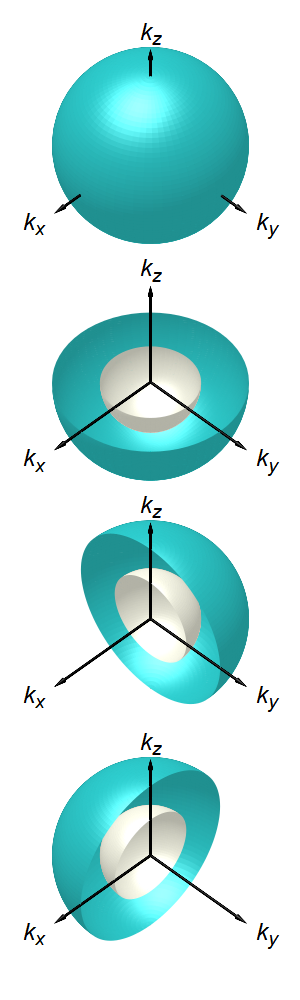}} & {\centering \includegraphics[width=0.14\textwidth]{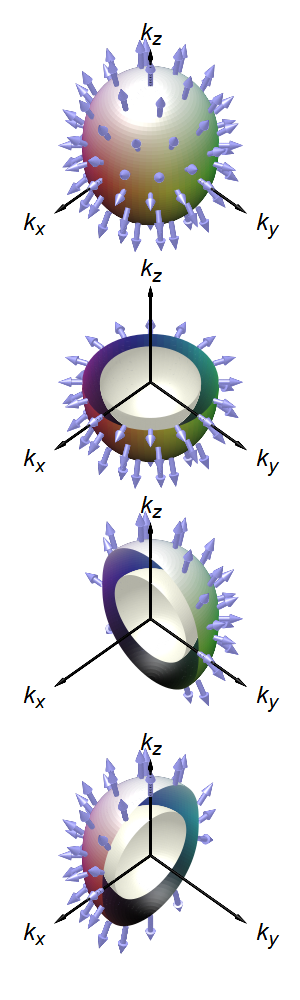}} & {\centering \includegraphics[width=0.14\textwidth]{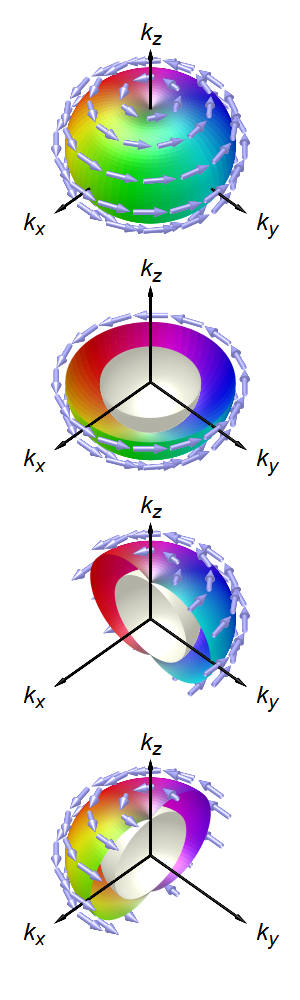}} & {\centering \includegraphics[width=0.14\textwidth]{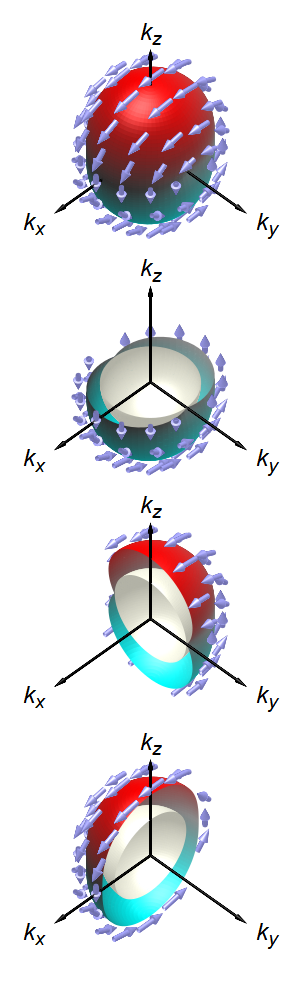}} & {\centering \includegraphics[width=0.14\textwidth]{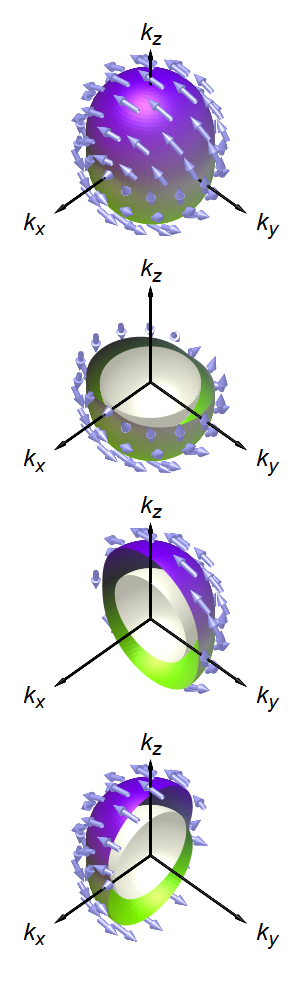}} \\
\bottomrule
\end{tabular}
\end{table}

One can easily notice in the figures that the $\Delta_{4x}$ and $\Delta_{4y}$ states have quite characteristic gap structures:
The $\Delta_{4x}$ state has a pair of point nodes ($|\bm{d}(\bm{k})| = 0$) along the $\pm k_y$ direction and $\Delta_{4y}$ has a pair of point-like gap minima along the $\pm k_x$ direction. 
This existence of a pair of gap node/minima violates the $C_3$ rotational symmetry of the crystal structure of \bs.
Thus, the $\Delta_{4x}$ and $\Delta_{4y}$ states are both gap-nematic SC states.
The nodes in $\Dx$ are protected by the mirror symmetry along the $a^\ast$-$c$ plane, whereas the nodes initially existing in $\Dy$ \revv{are} gapped out to become gap minima because there is no symmetry protecting the nodes~\revv{\cite{Fu2014.PhysRevB.90.100509}}.
In addition to the gap amplitude, the $\bm{d}$ vector structures of these $\Delta_4$ states also have nematic natures:
On the whole Fermi surface, the $\bm{d}$ vectors preferentially align along the $k_x$ direction in the $\Delta_{4x}$ state and along the $k_y$ direction in the $\Delta_{4y}$ state.
Reminding that a $\bm{d}$ vector is perpendicular to the Cooper-pair spin and the spin susceptibility should be small along the $\bm{d}$ vector, it is expected that the spin susceptibility of the $\Delta_4$ states exhibits two-fold behavior, with \revv{minima} along the $x$ direction for $\Delta_{4x}$ and $y$ for $\Delta_{4y}$~\cite{Hashimoto2013.JPhysSocJpn.82.044704}.
Thus, the $\Delta_{4x}$ and $\Delta_{4y}$ states are spin nematic \revv{as well}.

% Impurity robustness
\revv{We} should comment on the robustness of the proposed odd-parity states against non-magnetic impurity scattering.
Ordinarily, unconventional superconductivity is rather fragile against non-magnetic impurity scatterings, and such a strong reduction of superconductivity has been found in various non-$s$-wave superconductors~\cite{Sun1995.PhysRevB.51.6059, Mackenzie1998.PhysRevLett.80.161, Joo2004, Yonezawa2018.PhysRevB.97.014521}.
In doped \bs\ superconductors, because of the ion doping, impurity scattering is inevitably stronger than those of stoichiometric unconventional superconductors and can suppress the predicted odd-parity ($\Delta_2$, $\Delta_3$, and $\Delta_4$) states.
However, theories by Michaeli~\etal~\cite{Michaeli2012.PhysRevLett.109.187003} and Nagai~\etal~\cite{Nagai2015.PhysRevB.91.060502} proposed that these odd-parity states are rather robust against impurity scatterings because of strong spin-momentum locking in the normal-state Fermi surface.
Thus, odd-parity superconductivity itself as well as the nodal gap structure can be still stable in doped \bs.

\subsection{Early Experiments on the Superconducting State in Doped \bs}

% Before ~2013 (Cp, point-contact, STM, ....)

After the discovery of superconductivity in \cbs, various experiments were performed on the SC nature of this compound.
As a bulk probe, the temperature dependence of the electronic specific heat of \cbs\ was studied and was found to be different from the ordinary \revv{weak-coupling} BCS behavior~\cite{Kriener2011.PhysRevLett.106.127004}.
A theoretical calculation revealed that this $T$ dependence can be fitted well either with the $\Delta_2$ or $\Delta_4$-state models~\cite{Hashimoto2013.JPhysSocJpn.82.044704}.
\revv{Anomalous suppression of the superfluid density evaluated from the lower critical field $H\subm{c1}$ upon the change of the amount of Cu has been also attributed to the topological superconducting nature~\cite{Kriener2012.PhysRevB.86.180505}.}
The $T$ dependence of the upper critical field $\Hcc$ was investigated, and from the shape of $\Hcc(T)$ curve possibility of unconventional pairing was claimed~\cite{Bay2012.PhysRevLett.108.057001}.
However, in general, the shape of a $\Hcc(T)$ curve can vary merely due to changes in the Fermi-surface shape~\cite{Kita2004}.

Several surface-sensitive experiments, seeking for topological Majorana surface states, were also performed. 
The soft-point-contact spectroscopy has been the first experiment revealing the possible topological nature of the SC state~\cite{Sasaki2011.PhysRevLett.107.217001}.
A zero-bias peak in the differential conductivity as a function of the bias voltage was found and is attributed to topologically protected surface states.
Indeed, a theoretical calculation revealed that the observed conductivity is consistent with the odd-parity states~\cite{Yamakage2012.PhysRevB.85.180509}.
However, shortly later, a STM study on \cbs\ was performed and spectra resembling those of fully-gapped $s$-wave superconductivity was observed.
This apparent discrepancy may be due to the dimensionality of the system:
if the Fermi surface of \cbs\ is a quasi-two-dimensional cylinder rather than an ellipsoid (as indeed suggested by the ARPES and quantum oscillation experiment~\cite{Lahoud2013.PhysRevB.88.195107}), an STM spectrum on the $ab$ surface should be indistinguishable from the ordinary $s$-wave behavior. 
In contrast, a soft-point contact spectroscopy is expected to collect a certain average of conductivity of various directions, resulting in detecting the zero-bias anomaly in the $ab$-plane conductivity.
It was also claimed that STM spectra for $s$-wave superconductivity with an ellipsoidal Fermi surface should form a double peak structure, one originating from the coherence peak of the $s$-wave gap and the other from the surface state of the topological-insulator nature of \bs~\cite{Mizushima2014.PhysRevB.90.184516}.

Summarizing the experimental situation before the discovery of nematic superconductivity, there were evidence for topologically non-trivial superconductivity in \cbs\ but the debate was yet far from convergence.
It had been required to uncover more robust and reproducible properties evidencing for an interesting SC state.

%%%%%%%%%%%%%%%%%%%%%%%%%%%%%%%%%%%%%%%%%%
\section{Recent Experiments on Nematic Superconducting Behavior}
\label{sec:nematic-experiment}

\begin{table}[tbh]
\caption{Comparison of experimental reports on nematic superconductivity in doped \BS\ and related systems.}
\label{table:experimental_reports}
\centering
\tablesize{\footnotesize} %% You can specify the fontsize here, e.g.  \tablesize{\footnotesize}. If commented out \small will be used.
\renewcommand{\arraystretch}{1.4}
\begin{tabular}{ccccccc}
\toprule
\textbf{Material}	&	\textbf{Reference}	& \parbox{5em}{\centering\textbf{Growth method}$^{\mathrm{i}}$} & \parbox{5em}{\centering\textbf{Doping level \boldmath{$x$}}} & \textbf{Probe}	& \textbf{Large \boldmath{$\Hcc$}} & \parbox{6em}{\centering\textbf{Suggested state}}\\
\midrule
\multirow{3}{*}{\cbs}	& Matano 2016~\cite{Matano2016.NaturePhys.12.852}	& MG + ECI & 0.29-0.31 & NMR	& $y$ & $\Dx$ \\
\cline{2-7}
& Yonezawa 2017~\cite{Yonezawa2017.NaturePhys.13.123}	 & MG + ECI	& 0.3 & $C$ & $x$ & $\Dy$ \\
\cline{2-7}
& Tao 2018~\cite{Tao2018.PhysRevX.8.041024}	& MG + ECI & 0.31 & 	 STM	& -   & $\Dx$ \\ 
\cline{1-7}
\multirow{8}{*}{\sbs}	& Pan 2016~\cite{Pan2016.SciRep.6.28632}	& MG & 0.10, 0.15 & $\rho_{ab}$	& $x$ & \\
\cline{2-7}
& Nikitin 2016~\cite{Nikitin2016.PhysRevB.94.144516}	& MG & 0.15 & $\rho_{ab}$ in $P$	& $x$  & \\
\cline{2-7}
& \multirow{2}{*}{Du 2017~\cite{Du2017.SciChinaPhysMechAstron.60.037411}}	& \multirow{2}{*}{MG} & \multirow{2}{*}{NA} & \multirow{2}{*}{$\rho_{c}$}	& $x$ (\#1, \#2) & \\
\cline{6-7}
	&	&	&  & & $y$ (\#3)  & \\
\cline{2-7}
& Smylie 2018~\cite{Smylie2018.SciRep.8.7666}	& MG & 0.1 &  $\rho_{ab}$, $M$	& $x$  & \\
\cline{2-7}
& \multirow{2}{*}{Kuntsevich 2018~\cite{Kuntsevich2018.NewJPhys.20.103022}} &  \multirow{2}{*}{BG}	& \multirow{2}{*}{0.10-0.20}	& \multirow{2}{*}{$\rho_{ab}$}	& $x$ (some) & \\
\cline{6-7}
	&	&	&  & &  $y$ (others)  & \\
\cline{2-7}
& Willa 2018~\cite{Willa2018.PhysRevB.98.184509}	& MG & 0.1 & $C$	& $y$	 & \\
\cline{1-7}
\multirow{2}{*}{\nbs}	& Asaba 2017~\cite{Asaba2017.PhysRevX.7.011009}	& MG & NA &  torque & -  & \\
\cline{2-7}
& Shen 2017~\cite{Shen2017.npjQuantumMater.2.59}	& MG & 0.25 & $\rho_{ab}$, $M$ & y  & \\
\cline{1-7}
\rev{\cxpsbs} & \rev{Andersen 2018~\cite{Andersen2018.arXiv:1811.00805}}	& \rev{BG + ECI} & \rev{1.5} & \rev{$\rho_{ab}$, $\rho_c$, $C$} & \rev{$x$}  & \rev{$\Dx$} \\
\cline{1-7}
Bi\sub{2}Te\sub{3}/Fe(Se,Te) & Chen 2018~\cite{Chen2018.SciAdv.4.eeat1084} & MBE & - & STM & -  & $\Dy$ \\
\bottomrule
\multicolumn{7}{l}{${}^{\mathrm{i}}$ MG: melt growth, ECI: electrochemical intercalation, BG: \revv{Bridgman} method, MBE: molecular beam epitaxy}\\
%\multicolumn{7}{l}{${}^{\mathrm{ii}}$ $\Hccave \equiv [\Hcc(a)+\Hcc(a^\ast)]/2$}\\
%\multicolumn{7}{l}{${}^{\mathrm{iii}}$ $\Delta\Hcc \equiv \Hcc(a)-\Hcc(a^\ast)$}\\

\end{tabular}
\end{table}

In this section, we briefly review recent experimental findings on the nematic superconductivity in doped \bs\ superconductors.
As summarized in Table~\ref{table:experimental_reports}, nematic SC \revv{features have} been reported quite consistently in all of Cu-, Sr, and Nb-doped \bs\ superconductors as well as in related compounds, and \revv{have} been observed \revv{with} various bulk \revv{probes} such as \revv{nuclear magnetic resonance (NMR)} Knight shift, specific heat, resistivity and magnetization.
More recently, the STM technique has been successfully utilized to observe nematic features, adding microscopic evidence for the nematic superconductivity.

\subsection{Beginning of the Story: Nuclear Magnetic Resonance}

The pioneering work on the nematic superconductivity in doped \bs\ systems was performed by Matano~\etal, who investigated the spin susceptibility in the SC state of \cbs\ with the \revv{NMR} technique~\cite{Matano2016.NaturePhys.12.852}.
The spin susceptibility in the SC state is in general rather difficult to be measured, because of the strong Meissner screening. 
The NMR Knight shift is one of the few techniques that can measure the spin susceptibility directly, and has been utilized for various superconductors~\cite{Tou1996.PhysRevLett.77.1374, Ishida1998.Nature.396.658, Tien1989.PhysRevB.40.229, Shinagawa2007}.

\begin{figure}[tb]
\centering
\includegraphics[width=0.8\textwidth]{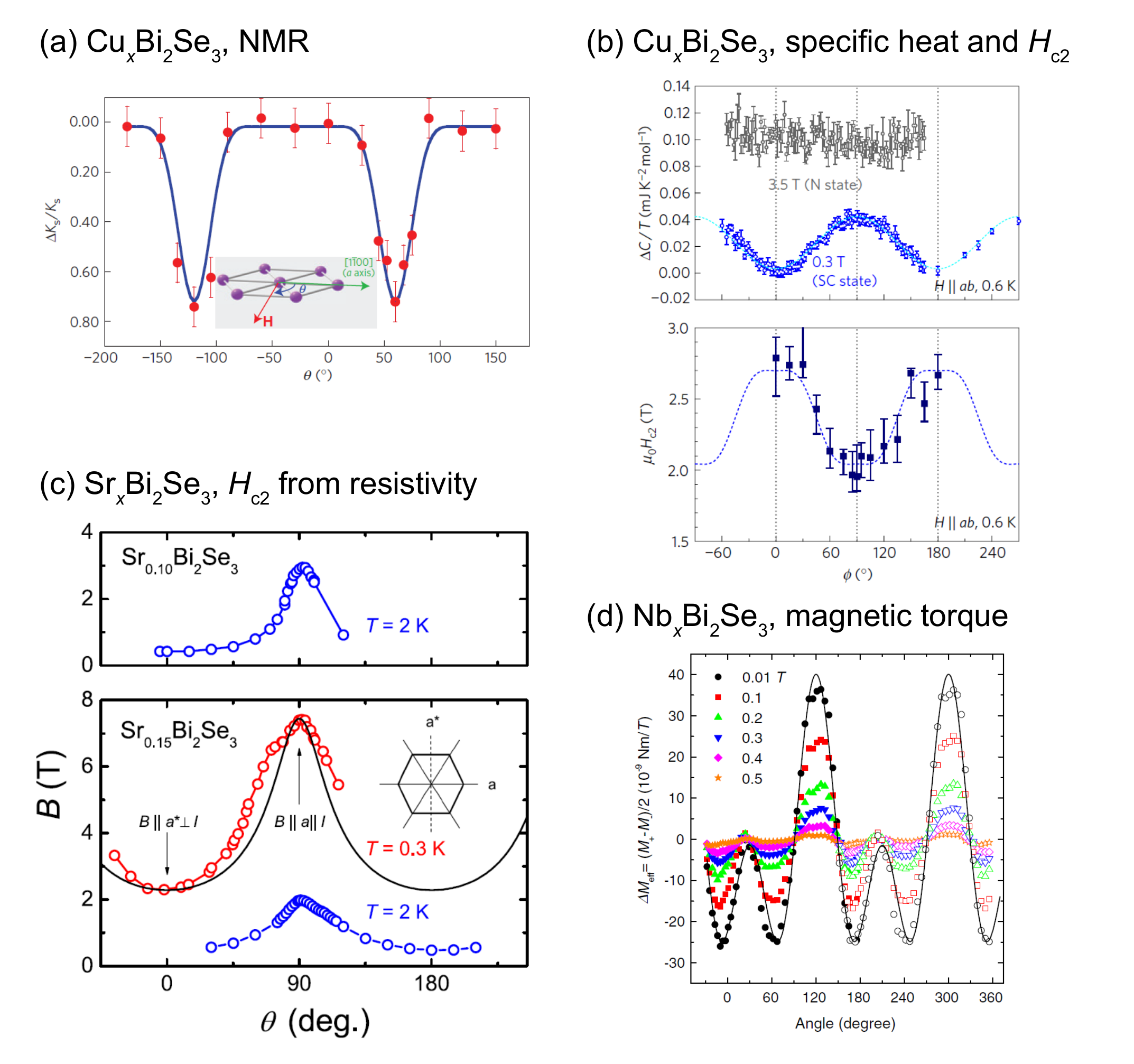}
\caption{Representative experiments on the nematic superconductivity in \abs.
(a) In-plane field-angle dependence of the NMR Knight shift of \cbs~\cite{Matano2016.NaturePhys.12.852}.
(b) In-plane field-angle dependence of the specific heat and $\Hcc$ of \cbs~\cite{Yonezawa2017.NaturePhys.13.123}.
(c) In-plane angular dependence of $\Hcc$ evaluated from magnetoresistance measurements on \sbs~\cite{Pan2016.SciRep.6.28632}.
(d) In-plane field-angle dependence of the irreversible component of the magnetic torque of \nbs~\cite{Asaba2017.PhysRevX.7.011009}.
We should be careful for the definition of the field angle: In (a) and (b), 0$^\circ$ corresponds to $H\parallel a$; whereas in (c) and (d), it corresponds to $H\parallel a^\ast$.
\revv{The panels (a) and (b) are respectively quated from Refs.~\cite{Matano2016.NaturePhys.12.852} and \cite{Yonezawa2017.NaturePhys.13.123} with a permission of Springer Nature; (c) and (d) are respectively from Refs.~\cite{Pan2016.SciRep.6.28632} and \cite{Asaba2017.PhysRevX.7.011009} under Creative Commons License.}
}
\label{fig:exper}
\end{figure}

Matano~\etal\ used ${}^{77}$Se nucleus (nuclear spin: 1/2) for NMR and investigate the spin susceptibility for various field directions within the $ab$ plane.
They used a set of four single crystals of \cbs\ with the Cu content $x$ of 0.29-0.31.
Usually the Knight shift contains the spin part and the other parts, and the latter was evaluated using NMR of non-doped \bs.
It was then found that the spin susceptibility decreases by nearly 80\% in the SC state for field directions parallel to one of the three equivalent crystalline $a$ axis, but does not change at all for other field directions, as shown in Fig.~\ref{fig:exper}(a). 
Overall, the spin susceptibility exhibit 180$^\circ$ periodicity as a function of in-plane field direction in spite of the trigonal crystalline lattice, evidencing that the spin part of the SC order parameter in \cbs\ is actually nematic.
This is not only the first clear observation of $SO(3)$ spin-rotational symmetry breaking in a SC state but also the first observation of the spin-nematic superconductivity in any known superconductors.

Results of this NMR work had been actually known in the community \revv{long} before the final publication in 2016, and stimulated \revv{subsequent} studies. 
In theories, Nagai~\etal\ already in 2012 pointed out that bulk superconducting properties such as thermal conductivity can exhibit rotational symmetry breaking if one of the $\Delta_4$ states is realized~\cite{Nagai2012.PhysRevB.86.094507}.
In 2014, Fu introduced the term ``nematic superconductivity'' in Ref.~\cite{Fu2014.PhysRevB.90.100509} and this suitable name seems to have contributed significantly to the expansion of the research field.

\subsection{Pioneering Reports of Bulk Properties}

In early 2016, three subsequent works reporting the nematic bulk SC nature in doped \bs\ were independently and almost simultaneously submitted to the arXiv server, and were later published in 2016-2017~\cite{Yonezawa2017.NaturePhys.13.123, Pan2016.SciRep.6.28632, Asaba2017.PhysRevX.7.011009}.
As briefly reviewed below, these three works by different groups consistently revealed \revv{the} nematic nature for different dopants (Cu, Sr, and Nb) and using different experimental probes, demonstrating that the nematic superconductivity is a robust and ubiquitous feature in doped \bs\ superconductors.

\revv{The present author and coworkers} performed specific-heat measurements of \cbs\ single crystals under precise two-axis field-direction control~\cite{Yonezawa2017.NaturePhys.13.123}.
Since the electronic specific heat is quite small in doped \bs\ because of low carrier concentration and weak electron correlation, it was necessary to measure the specific heat with high resolution.
To \revv{achieve} this goal, they built a small and low-background calorimeter utilizing the AC technique~\cite{Sullivan1968}, which has the highest resolution among the standard heat-capacity measurement techniques. 
To apply magnetic fields, they used a vector magnet system~\cite{Deguchi2004RSI}, allowing them to perform two-axis field-direction control.
With this system, together with a careful field-alignment process, field misalignment effects were minimized.
It was then found that the specific heat as a function of the in-plane field angle exhibit a two-fold symmetric behavior (Fig.~\ref{fig:exper}(b)), clearly breaking the lattice $C_3$ rotational symmetry.
This nematicity in a bulk thermodynamic quantity is only possible if the SC gap amplitude has a nematic nature.
Thus, this specific-heat result provides the first thermodynamic evidence for the gap-nematic superconductivity.
From a comparison between the observed specific-heat oscillation and theoretical calculation, it was concluded that the $\Dy$ state is realized.
In addition, the upper critical field $\Hcc$ was also found to exhibit two-fold behavior with in-plane anisotropy of 20\% as shown in Fig.~\ref{fig:exper}(b), providing additional evidence for the nematic superconductivity.

Pan \etal\ measured the in-plane resistivity of \sbs\ under in-plane magnetic fields, and observed in-plane $\Hcc$ anisotropy of around 400\%~\cite{Pan2016.SciRep.6.28632} as plotted in Fig.~\ref{fig:exper}(c).
In this experiment, strictly speaking, the applied electric current explicitly breaks the in-plane rotational symmetry.
Nevertheless, the observed anisotropy is huge and cannot be explained by the anisotropy due to the applied current. 
Indeed, the absence of the role of the electric current direction on the SC nematicity was later confirmed by $c$-axis resistivity measurements~\cite{Du2017.SciChinaPhysMechAstron.60.037411} as well as by  in-plane resistivity measurements \revv{upon} varying the current direction~\cite{Kuntsevich2018.NewJPhys.20.103022}.
This work has another important aspect for demonstrating that a simple technique such as resistivity can probe the nematicity.
Shortly later, Nikitin~\etal\ (the same group as Pan \etal) reported that the nematic SC feature is robust even under hydrostatic pressure~\cite{Nikitin2016.PhysRevB.94.144516}.

Asaba~\etal\ investigated \nbs\ single crystals by means of torque magnetometry under various in-plane field directions~\cite{Asaba2017.PhysRevX.7.011009}.
They studied the size of the hysteresis between the field-up and down sweep torque signals.
They found that this hysteresis size exhibits clear breaking of the expected six-fold symmetric behavior as a function of \revv{the} in-plane field angle (Fig.~\ref{fig:exper}(d)).
\revv{The} hysteresis size is \revv{actually} not a thermodynamic quantity but is rather related to the vortex pinning and the critical current \revv{density} of the sample.
Thus, it is not very straightforward how this quantity is related to the nematic SC gap.
Nevertheless, it is also difficult to come up with other extrinsic origins on this nematic hysteresis.
The mechanism of this interesting appearance of nematicity in the hysteresis should be clarified in future.
 
\subsection{Recent Reports}

More recently, many other groups reported nematic superconductivity in doped \bs. 
In particular, many works have been performed for Sr-doped \bs. 
Du~\etal\ measured $c$-axis resistance under magnetic field to avoid the symmetry breaking due to the external current, and found that the nematicity is still there; thus the external current is not the origin of \revv{the observed} two-fold behavior in the in-plane resistivity~\cite{Du2017.SciChinaPhysMechAstron.60.037411}.
More interestingly, they investigated the sample dependence of the nematic behavior, and found that the anisotropy of the upper critical field, \revv{depicting} the nematicity, is actually sample dependent: among the three samples investigated, two \revv{have a} large $\Hcc$ for the $x$ direction but the other has a large $\Hcc$ for the $y$ direction.
Such a sample dependence \revv{implies} that the ground state ($\Delta_{4x}$ or $\Delta_{4y}$) is \revv{also sample dependent}.
This issue will be discussed in more detail in the next section. 
Smylie~\etal\ investigated the in-plane field-angle dependence of resistivity and magnetization~\cite{Smylie2018.SciRep.8.7666}. 
The observed two-fold behavior in the magnetization provides the first thermodynamic evidence of the gap-nematic superconductivity in the Sr-doped compound.
They also investigated the crystal structure using X-ray diffraction, and concluded that there is no detectable crystalline distortion in their sample. 
Kuntsevich~\etal\ reported in-plane resistivity anisotropy using samples grown with the \revv{Bridgman} method~\cite{Kuntsevich2018.NewJPhys.20.103022}. 
They cut two kinds of samples from the same batch: ones cut along the $a$ axis and the others cut along the $a^\ast$ axis, to check the current-direction dependence of the nematicity. 
It was found that the nematicity is independent of the current direction.
But the nematicity is actually dependent on batches, confirming the sample-dependent nematicity reported in Ref.~\cite{Du2017.SciChinaPhysMechAstron.60.037411}. 
They also reported a tiny crystal deformation as well as two-fold resistivity anisotropy even in the normal state.
This possible ``normal-state nematicity'' will be discussed in Sec.~\ref{subsec:normal-state-nematicity}.
Willa~\etal\ succeeded in measuring the specific heat of \sbs~\cite{Willa2018.PhysRevB.98.184509}.
Calorimetry of \sbs\ is more challenging than that of \cbs, because the carrier density and resulting electronic specific heat tends to be much lower in \sbs~\cite{Liu2015.JAmChemSoc.137.10512, Shruti2015.PhysRevB.92.020506}.
Nevertheless, they used a micro-structured calorimeter to achieve sensitivity high enough for a \sbs\ single crystal and resolved a specific-heat jump of around 0.25~mJ/K\sps{2}mol, which is only a fraction of that observed in \cbs~\cite{Kriener2011.PhysRevLett.106.127004, Yonezawa2017.NaturePhys.13.123}.
The two-fold specific-heat behavior was then observed, which was attributed to large $\Hcc$ anisotropy rather than the gap anisotropy.

For Nb-doped \bs, Shen~\etal\ reported in-plane field-angle dependence of magnetization and resistivity~\cite{Shen2017.npjQuantumMater.2.59}.
They found nematicity in both quantities, providing the first thermodynamic and transport evidence for the gap-nematic superconductivity in \nbs.
\revv{We} should also mention that a penetration-depth measurement was performed \revv{on} \nbs~\cite{Smylie2017.PhysRevB.96.115145}.
The penetration depth was found to exhibit a $T^2$-temperature dependence down to $T/\Tc \sim 0.12$.
This behavior is consistent with the existence of point-nodes or point-like very small gap minima. 
This result provides an indirect support for the nematic $\Delta_{4}$ states. 
It was also found that the $T^2$ dependence is robust against the increase of \revv{the} impurity concentration. 
This robustness is consistent with theoretical proposals~\cite{Michaeli2012.PhysRevLett.109.187003, Nagai2015.PhysRevB.91.060502}.

\rev{
Quite recently, Andersen~\etal\ reported that Cu-intercalated \psbs, which is a ``naturally-made heterostructure'' of PbSe and \bs\ layers and exhibits superconductivity below $\Tc\sim 2.5$~K after intercalation~\cite{Sasaki2014.PhysRevB.90.220504}, exhibits two-fold anisotropy in the SC state via resistivity and specific-heat measurements~\cite{Andersen2018.arXiv:1811.00805}. 
Strictly speaking, the global crystal symmetry of this compound is orthorhombic and does not have three-fold rotational symmetry due to the neighboring PbSe layer. 
Nevertheless, the \bs\ layer of this compound, as well as its electronic structure, almost preserves the three-fold symmetry~\cite{Sasaki2014.PhysRevB.90.220504, Nakayama2015.PhysRevB.92.100508}.
Therefore, the observed two-fold anisotropy in the SC state is most likely due to \revv{a} nematic SC gap, rather than the conventional origin such as Fermi-velocity anisotropy.
It is worth commenting that \cxpsbs\ substantially differs from \abs\ in various aspects: \cxpsbs\ has a highly two-dimensional electronic structure because of separation of the conductive \bs\ layers by insulating PbSe layers, and exhibit line-nodal SC behavior~\cite{Sasaki2014.PhysRevB.90.220504}.
This observation implies that the nematic superconductivity is quite robust irrespective of the dimensionality and gap structure of the system, thus providing important information on the origin and nature of nematic superconductivity.
}

\subsection{Direct Visualization}

Recently, direct visualization of nematic superconductivity by STM has been reported. 
Chen~\etal\ investigated \bt\ thin films (with typical thickness of 2QL) grown on a FeSe$_{0.55}$Te$_{0.45}$ single-crystal substrate via molecular-beam epitaxy (MBE)~\cite{Chen2018.SciAdv.4.eeat1084}.
This system is a bit different from doped \bs, because the \bt\ thin film exhibits superconductivity due to the proximity effect from the substrate, whereas doped \bs\ exhibits bulk superconductivity.
On the other hand, the surface of \bt/FeSe$_{0.55}$Te$_{0.45}$ can be cleaner than \abs\ because no ion doping is necessary to induce superconductivity, being more suited for STM investigations.
In the quasiparticle interference spectra of \bt/FeSe$_{0.55}$Te$_{0.45}$ at zero field, the quasiparticle excitation in the intermediate energy range below the SC gap was found to exhibit two-fold anisotropy.
The excitation is stronger along the $\pm k_x$ direction, indicating that the SC gap is smaller for this field direction, thus suggesting the $\Delta_{4y}$ state.
In addition, they found that the magnetic vortices under $H\parallel c$ have an ellipsoidal shape elongating along the $x$ ($a$) direction, although a nearly isotropic vortex shape is expected for the trigonal crystal structure.
Such a vortex core shape provides microscopic evidence for two-fold anisotropy in the penetration depth and in the coherence length.

Independently, an STM study of \cbs\ was reported by Tao~\etal~\cite{Tao2018.PhysRevX.8.041024}. 
In spite of difficulties in finding good surfaces for STM, they succeeded in obtaining STM spectra under magnetic field.
Under $c$-axis field, they observed ellipsoidal vortex cores elongating along the $y$ ($a^\ast$) direction, breaking the $C_3$ rotational symmetry.
Moreover, they investigated the dependence of the gap amplitude on the in-plane field directions. 
They directly observed that the SC gap amplitude exhibits a two-fold field-angle dependence.
The gap is \revv{larger} for $H\parallel y$ than for $H\parallel x$ at 0.5~T, in favor of the $\Dx$ state.
At higher fields, the large-gap axis rotates by $\sim 20^\circ$.
The origin of this rotation is not clear yet.

These STM studies are significant in providing clear microscopic evidence for gap-nematic superconductivity.
Also, they are the first observations of nematicity in the absence of \revv{an} in-plane magnetic field; it is now confirmed that the nematic superconductivity is the ground state even at zero in-plane field.

%%%%%%%%%%%%%%%%%%%%%%%%%%%%%%%%%%%%%%%%%%
\section{Known Issues}
\label{sec:issues}

Although the nematic feature in the SC state has been consistently and reproducibly observed in \abs\ as well as in a related compound, there are several controversial or unresolved issues.
These issues are \revv{addressed} in this section.

\subsection{Which of $\Delta_{4x}$ or $\Delta_{4y}$ are Realized?}

One of the most important but puzzling issue is which of $\Delta_{4x}$ or $\Delta_{4y}$ are realized in actual samples.
As listed in Table~\ref{table:experimental_reports}, the NMR~\cite{Matano2016.NaturePhys.12.852} and STM~\cite{Tao2018.PhysRevX.8.041024} studies on \cbs\ suggested the $\Delta_{4x}$ state but the specific heat study~\cite{Yonezawa2017.NaturePhys.13.123} on the same compound suggested the $\Delta_{4y}$ state.
Moreover, the $\Hcc$ anisotropy, an indicator of the nematic direction, significantly varies: some reports \revv{suggest a} large $\Hcc$ for $H\parallel x$ and others for $H\parallel y$.
Works on multiple samples indicate that the $\Hcc$ anisotropy is actually sample dependent~\cite{Du2017.SciChinaPhysMechAstron.60.037411, Kuntsevich2018.NewJPhys.20.103022}.
Thus, the variation in the $\Hcc$ anisotropy is not an artifact but is an intrinsic property of \abs.

\begin{figure}[tb]
\centering
\includegraphics[width=0.9\textwidth]{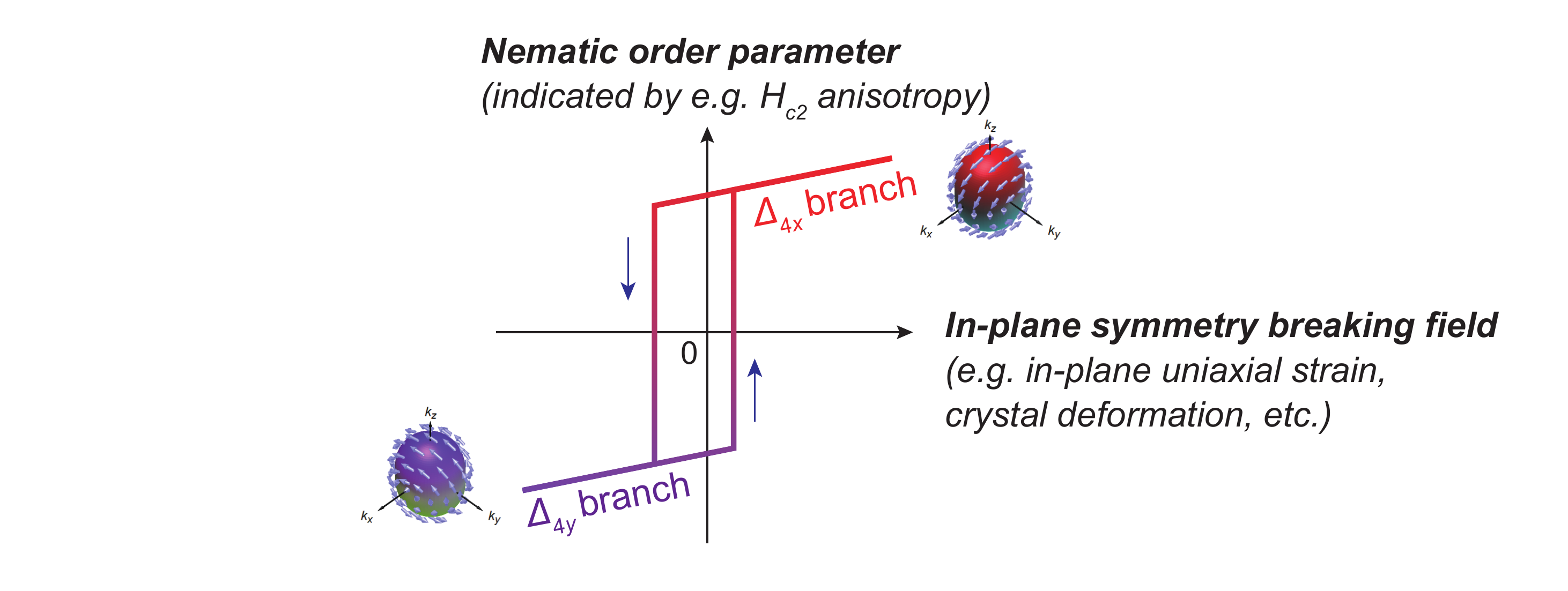}
\caption{Schematic figure on the coupling between the nematic order parameter and the in-plane symmetry breaking field.}
\label{fig:branches}
\end{figure}

This fact very likely suggests that the $\Dx$ and $\Dy$ states are  nearly degenerate and are chosen depending on the in-plane symmetry breaking fields. 
Such a situation is described in Fig.~\ref{fig:branches}: the nematic SC order parameter, which is reflected in, e.g., in-plane $\Hcc$ anisotropy, exhibits multiple branches as a function of the in-plane symmetry breaking field, such as in-plane uniaxial strain~\cite{Venderbos2016.PhysRevB.94.094522}, crystal deformation, or some arrangement of doped ions. 
This coupling between the order parameter and the symmetry-breaking field resembles the coupling between ferromagnetism and external magnetic fields.

The \revv{schematic} model in Fig.~\ref{fig:branches} explains the reason of the variation in the large $\Hcc$ direction: for some samples $\Dx$ is favored because of a tiny but positive symmetry breaking field and for others $\Dy$ is realized due to negative field.
This also explains the absence of random switching of nematicity orientation upon different coolings:
So far there are no reports that the nematic behavior is altered by different cooling histories across $\Tc$.
Thus all the samples seem to have finite symmetry-breaking fields that pin the nematicity.

\subsection{Normal-State and Superconducting-State Nematicities}
\label{subsec:normal-state-nematicity}

This model also explain the relation between the SC nematicity and the possible lattice distortion.
Because it is quite important whether the trigonal lattice symmetry is preserved even after the doping, the existence of lattice symmetry breaking has been investigated.
Until recently, there has been no positive evidence for the lattice distortion as well as for nematic behavior in the normal state~\cite{Matano2016.NaturePhys.12.852, Yonezawa2017.NaturePhys.13.123, Smylie2018.SciRep.8.7666}.
Recently, Kuntsevich~\etal\ reported that their samples grown using the \revv{Bridgman} method (in which crystals tend to grow along one direction) have a tiny lattice distortion (0.02\% orthorhombic distortion and $0.005^\circ$ $c$-axis inclination)~\cite{Kuntsevich2018.NewJPhys.20.103022}.
These samples exhibit in-plane $\Hcc$ anisotropies ranging 300-800\%.
Such large anisotropy cannot be explained at all by the electronic anisotropy caused by the observed tiny lattice distortion.
Therefore, SC nematicity is mostly caused by the Cooper-pair formation and the concept of nematic superconductivity should be still valid in the presence of explicit breaking of the trigonal symmetry.
These samples with relatively large SC anisotropy and detectable lattice distortion are probably located in the region far from the origin in the schematic \revv{in Fig.~\ref{fig:branches}}.

\revv{The} nematic SC features observed in \cxpsbs~\cite{Andersen2018.arXiv:1811.00805} and Bi\sub{2}Te\sub{3}/Fe(Se,Te)~\cite{Chen2018.SciAdv.4.eeat1084} are also regarded as nematic superconductivity in explicit symmetry-breaking fields. 
In these systems, the global crystal structures do not possess trigonal rotational symmetry due to the neighboring non-trigonal layers.
Thus, there are finite symmetry breaking fields in these compounds.
Still, anisotropies in their normal states are rather small and are not sufficient to explain the sizable two-fold anisotropies in their SC states.

\subsection{Nematic Domains}

If the symmetry-breaking field is rather weak, a formation of multiple nematic domains is expected.
In the specific-heat study \revv{in Ref.~\cite{Yonezawa2017.NaturePhys.13.123}}, one sample with possibe nematic domains was reported (See the Supplementary Information of Ref.~\cite{Yonezawa2017.NaturePhys.13.123}).
This sample exhibits very weak $\Hcc$ anisotropy, as well as weak and distorted specific-heat oscillation as a function of in-plane magnetic field direction, in contrast to the sample mainly focused on in Ref.~\cite{Yonezawa2017.NaturePhys.13.123}. 
Angular magnetoresistance (AMR) is more sensitive to the existence of domains.
Indeed, complicated structures in AMR curves attributable to domains were observed in Refs.~\cite{Du2017.SciChinaPhysMechAstron.60.037411, Kuntsevich2018.NewJPhys.20.103022}.

However, in most of the other studies, samples seem to be in single-domain states.
For example, in the NMR study~\cite{Matano2016.NaturePhys.12.852}, multiple domains would result in multiple sets of dips in the Knight shift as a function of in-plane field angle. 
In the actual experiment, only one set of dips was observed, suggesting a single-domain sample.
Such dominance of single-domain samples indicate that the in-plane symmetry breaking field is in most cases strong enough to avoid the formation of multiple domains.

A nematic domain wall, if \revv{existing}, is a fascinating object, forming naturally a junction between topological pairing states.
This may host novel Majorana quasiparticles and may be controllable by \revv{an} external symmetry breaking field.
Investigation of detailed order-parameter structures near the domain wall would be quite interesting.

\subsection{Possible Nematic Superconductivity in Other Systems}

In principle, gap-nematic superconductivity can occur in any type of superconductivity, even in ordinary $s$-wave superconductivity.
Nevertheless, a straightforward way to realize nematic superconductivity is to bring a multi-component superconductor (i.e., a superconductor with a multi-dimensional irreducible representation) and to stabilize one of the SC-order-parameter components.
In such multi-component superconductivity, each component \revv{is} in most cases nematic.
But usually the components form a complex linear combination to satisfy the rotational symmetry of the underlying lattice. 
For example, in the chiral $p_x \pm i p_y$-wave superconductivity \revv{on} a tetragonal lattice \revv{(Fig.~\ref{fig:nematic-schematic}(b))}, each $p_x$ or $p_y$ component breaks the tetragonal symmetry, thus possessing the nematic nature. 
However, they form the complex (chiral) combination $p_x \pm i p_y$ to satisfy the tetragonal symmetry except in the SC phase degree of freedom.
If, however, the formation of such a chiral state is unfavored,
one of the nematic components can be stabilized.

Actually, in doped \bs, the predicted nematic $\Delta_{4x}$ and $\Delta_{4y}$ states both belong to the two-dimensional $E_u$ representation, as described in Table~\ref{table:SC-states}. 
The two components thus in principle can form complex (chiral) combinations.
However, in the present case, the strong spin-momentum locking forces a non-unitary SC state to emerge when the chiral superconductivity is realized~\cite{Venderbos2016.PhysRevB.94.180504}.
Notice that the $\bm{d}$ vector of the chiral state $\bm{d}\subm{chiral} = \bm{d}_{4x} \pm i\bm{d}_{4y} \sim (k_z \pm i\varepsilon k_x, \mp ik_z, \lambda(k_x \pm i k_y))$ has a complex spin component:
See, for example, along the $k_z$ axis, $\bm{d}\subm{chiral} \sim \pm k_z(\hat{\bm{x}} - i\hat{\bm{y}})$ is non-unitary ($\bm{d}\times\bm{d}^{\ast} \ne 0$).
Generally, non-unitary states have spin-dependent excitation gaps and usually one spin \revv{component has a} significantly smaller gap than the other. 
Thus, non-unitary states are expected to have smaller condensation energies than the ordinary unitary SC states.
This inevitable formation of non-unitary gaps in chiral states prevents the formation of complex linear combination of $\bm{d}_{4x}$ and $\bm{d}_{4y}$ and favors the realization of single-component nematic superconductivity in this system.

From the discussion above, a multi-component superconductor is a promising platform \revv{to probe} nematic superconductivity.
Indeed, in UPt\sub{3} with a trigonal crystal structure $P\bar{3}m1$~\cite{Walko2001.PhysRevB.63.054522}, multi-component superconductivity is believed to be realized~\cite{Izawa2014.JPhysSocJpn.83.061013} with one of the components relatively stabilized by the short-ranged antiferromagnetic ordering~\cite{Aeppli1988.PhysRevLett.60.615}.
In the in-plane field-angle dependence of the thermal conductivity, two-fold symmetric behavior was observed in the C phase~\cite{Machida2012.PhysRevLett.108.157002, Izawa2014.JPhysSocJpn.83.061013}.
This phenomenon can be considered as a consequence of nematic superconductivity with a finite in-plane symmetry-breaking field (see Fig.~\ref{fig:branches}).
In PrOs\sub{4}Sb\sub{12} with a cubic structure $Im\bar{3}$, similar two-fold behavior in the field-angle-dependent heat transport was observed in a part of the $H$-$T$ phase diagram~\cite{Izawa2003.PhysRevLett.90.117001}. 
Strictly speaking, two-fold symmetry in the magneto-transport is allowed in the space group $Im\bar{3}$.
Nevertheless, it seems difficult to explain the clear two-fold anisotropy just due to the electronic-state anisotropy, and gap nematicity due to a multi-component order parameter may be playing an important role.
On the other hand, two-fold behavior has not been observed in the specific heat of both superconductors~\cite{Sakakibara2007.JPhysSocJpn.76.051004.review, Kittaka2013.JPhysSocJpn.82.024707}.
Thus, the nematic feature of these compounds, if \revv{existing}, is subtle compared with those observed in doped \bs\ superconductors. 
Theoretically, several other multi-comonent-superconductor candidates such as U$_{1-x}$Th$_x$Be\sub{13}~\cite{Machida2018.JPhysSocJpn.87.033703} and half-Heusler compounds~\cite{Roy2017.arXiv:1708.07825, Venderbos2018.PhysRevX.8.011029} were proposed to exhibit nematic superconductivity, and experimental verification is \revv{strongly called for}.

\revv{We} also comment on another leading candidate of multi-component superconductivity, \sro, seemingly exhibiting quasi-two-dimensional chiral $p_x\pm ip_y$-wave superconductivity. 
Theoretically, it has been predicted \revv{that} a non-chiral single-component $p_x$ (or $p_y$) state~\cite{Agterberg1998.PhysRevLett.80.5184} \revv{should be realized} under in-plane magnetic field or in-plane uniaxial strain.
Such non-chiral states may be regarded as a symmetry-breaking-field-induced nematic state, or a ``meta-nematic'' state in an analogy to the metamagnetic transition (field-induced ferromagnetism) in a paramagnet.
Although clear experimental observation has not been achieved yet~\cite{Yonezawa2013.PhysRevLett.110.077003, Yonezawa2014.JPhysSocJpn.83.083706, Hicks2014.Science.344.283, Steppke2017.Science.355.eaaf9398}, it is worth investigating seriously nematic features of \sro\ under in-plane symmetry breaking fields.
\revv{Lastly, we mention that nematic superconductivity in \sro\ at ambient condition is theoretically proposed very recently in Ref.~\cite{Huang2018.PhysRevLett.121.157002}.}

%%%%%%%%%%%%%%%%%%%%%%%%%%%%%%%%%%%%%%%%%%
\section{Summary and Perspectives}
\label{sec:conclusion}

To summarize, \revv{we} have reviewed recent researches on nematic superconductivity in doped \bs\ topological superconductors \rev{and in related compounds}. 
Two-fold symmetric behavior in many quantities, breaking the trigonal symmetry of the underlying lattice, has been reported by more than ten groups, with excellent reproducibility. 
These experimental works demonstrate that the nematic superconductivity is a common and robust feature among the \abs\ family.
In addition, observation of nematic gap \revv{structures} in turn provide bulk evidence for topological superconductivity in this family~\revv{\cite{Fu2010.PhysRevLett.105.097001}}.
However, there are several issues unresolved; in particular, apparent inconsistency in the nematic direction and its relation to the possibly existing in-plane symmetry-breaking field are most important subjects to be investigated next.
Also, predicted novel phenomena originating from nematic superconductivity, such as the superconductivity-fluctuation-induced nematic order above $\Tc$~\cite{Hecker2018.npjQuantumMater.3.26}, the chiral Higgs mode in the electromagnetic response~\cite{Uematsu2018.arXiv:1809.06989}\revv{, spin polarization of Majorana quasiparticles in a vortex core~\cite{Nagai2014.JPhysSocJpn.83.064703}, and nematic Skyrmion texture near half-quantum vortices~\cite{Zyuzin2017.PhysRevLett.119.167001},} would be worth seeking for.

The nematic SC state in \abs\ is qualitatively unique compared with other nematic systems in liquid crystals or normal-state electron systems:
The nematic superconductivity is realized by a macroscopically-coherent quantum-mechanical wavefunction, accompanied by an odd-parity nature, non-trivial topology, and an active spin degree of freedom. 
\revv{We} believe that this new class of nematic states stimulates further researches both in the fields of nematic liquids and of unconventional topological superconductivity.

%**** Something nice should be mentioned *****

%%%%%%%%%%%%%%%%%%%%%%%%%%%%%%%%%%%%%%%%%%
\vspace{6pt} 

%%%%%%%%%%%%%%%%%%%%%%%%%%%%%%%%%%%%%%%%%%
%% optional
%\supplementary{The following are available online at \linksupplementary{s1}, Figure S1: title, Table S1: title, Video S1: title.}

% Only for the journal Methods and Protocols:
% If you wish to submit a video article, please do so with any other supplementary material.
% \supplementary{The following are available at \linksupplementary, Figure S1: title, Table S1: title, Video S1: title. A supporting video article is available at doi: link.}

%%%%%%%%%%%%%%%%%%%%%%%%%%%%%%%%%%%%%%%%%%
\authorcontributions{Writing: S.Y.}

%%%%%%%%%%%%%%%%%%%%%%%%%%%%%%%%%%%%%%%%%%
\funding{This research was funded by the Japanese Society for Promotion of Science (JSPS) grant number KAKENHI JP15H05851, JP15H05852, JP15K21717, JP26287078, and JP17H04848}

%%%%%%%%%%%%%%%%%%%%%%%%%%%%%%%%%%%%%%%%%%
\acknowledgments{The author acknowledges collaboration with K.~Tajiri, S.~Nakata, I.~Kostylev, Y.~Maeno at Kyoto University, Z.~Wang and Y.~Ando at the University of K\"{o}ln, K.~Segawa at Kyoto Sangyo University, and Y.~Nagai at Japan Atomic Energy Agency. The author also thanks \revv{M.~Kriener for valuable comments to improve the manuscript}, and \rev{K.~Ishida, Y.~Matsuda, T.~Mizushima, M.~Sato, L.~Fu, K.~Matano, G.-q.~Zheng, and  D.~Agterberg} for valuable discussion and support.}

%%%%%%%%%%%%%%%%%%%%%%%%%%%%%%%%%%%%%%%%%%
\conflictsofinterest{The authors declare no conflict of interest.} 

%%%%%%%%%%%%%%%%%%%%%%%%%%%%%%%%%%%%%%%%%%
%% optional
\abbreviations{The following abbreviations are used in this manuscript:\\

\noindent 
\begin{tabular}{@{}ll}
AMR & Angular magnetoresistance\\
ARPES & Angle-resolved photoemission spectroscopy\\
BCS & Bardeen-Cooper-Schrieffer\\
BG  & \revv{Bridgman} method \\
BW  & Balian-Werthamer \\
ECI & Electrochemical intercalation \\
MBE & Molecular-beam epitaxy \\
MG  & Melt growth\\
NMR & Nuclear magnetic resonance\\
QL & Quintuple layer\\
SC & Superconducting\\
STM & Scanning tunneling microscope\\
vdW & van der Waals
\end{tabular}}

%=====================================
% References, variant B: external bibliography
%=====================================
\externalbibliography{yes}

\bibliography{%
../../bib/string,%
../CuBiSe_review,%
../../bib/3He.bib,%
../../bib/TMTSF,%
../../bib/Fe-Pn,%
../../bib/textbook,%
../../bib/FFLO,%
../../bib/CeCoIn5,%
../../bib/heavy-Fermion,%
../../bib/superconductors,%
../../bib/SC,%
../../bib/Sr2RuO4,%
../../bib/Sr3Ru2O7,%
../../bib/high-Tc,%
../../bib/others,%
../../bib/measurement_technique,%
../../bib/CuxBi2Se3_v2,%
../../bib/Topological_insulator,%
../../bib/Topological_superconductor%
}

%%%%%%%%%%%%%%%%%%%%%%%%%%%%%%%%%%%%%%%%%%
%% optional
%\sampleavailability{Samples of the compounds ...... are available from the authors.}

%%%%%%%%%%%%%%%%%%%%%%%%%%%%%%%%%%%%%%%%%%
\end{document}